# CB-HVTNet: A channel-boosted hybrid vision transformer network for lymphocyte assessment in histopathological images

***Momina Liaqat Ali[1], Zunaira Rauf[1, 2], Asifullah Khan[1, 2, 3*], Anabia Sohail[4], Rafi Ullah[5], Jeonghwan Gwak [6*]***

[1]Pattern Recognition Lab, Department of Computer & Information Sciences, Pakistan Institute of Engineering & Applied Sciences, Nilore, Islamabad 45650, Pakistan
[2]PIEAS Artificial Intelligence Center (PAIC), Pakistan Institute of Engineering & Applied Sciences, Nilore, Islamabad 45650, Pakistan
[3]Center for Mathematical Sciences, Pakistan Institute of Engineering & Applied Sciences, Nilore, Islamabad 45650, Pakistan
[4]Department of Computer Science, Faculty of Computing and Artificial Intelligence, Air University, E-9, Islamabad, Pakistan
[5]Department of Computer and Information Sciences, Universiti Teknologi PETRONAS (UTP), 31750, Perak, Malaysia
[6]Department of Software, Korea National University of Transportation, Chungju 27469, Republic of Korea
**Corresponding Authors:** *Asifullah Khan, asif@pieas.edu.pk, *****Jeonghwan Gwak, jgwak@ut.ac.kr

## Abstract

Transformers, due to their ability to learn long-range dependencies, have overcome the shortcomings of convolutional neural networks (CNNs) for global perspective learning. However, their multi-head attention module only captures global-level feature representations, which is insufficient for medical images. To address this issue, we propose a Channel Boosted Hybrid Vision Transformer (CB-HVT) that uses transfer learning to generate boosted channels and employs both transformers and CNNs to analyse lymphocytes in histopathological images. The proposed CB-HVT comprises five modules to effectively identify lymphocytes. Its (i) channel generation module, uses the idea of channel boosting through transfer learning to extract diverse channels from different auxiliary learners. These boosted channels are first concatenated and ranked using an attention mechanism in the (ii) channel exploitation module. A fusion block is then utilized in the (iii) channel merging module for a gradual and systematic merging of the diverse boosted channels to improve the network's learning representations. The CB-HVT also employs a proposal network in its (iv) region aware module followed by its (v) detection and segmentation head to accurately identify and distinguish objects, even in the regions with crowded presence and artifacts. We evaluated the proposed CB-HVT on two publicly available datasets for lymphocyte assessment in histopathological images. The results demonstrate that CB-HVT has a good generalization on unseen data therefore, it can serve as a valuable tool for pathologists.

# 1. Introduction

Recent advancements in Convolutional Neural Networks (CNNs) have revolutionized computer vision and facilitated the development of computer-aided diagnostic systems (J. Gao et al., 2019). CNN-based automated systems have been developed for various tasks, including tumor classification, nuclei segmentation, COVID detection, and cancer analysis, because of their ability to learn discriminant features automatically from images (Ke et al., 2023; S. H. Khan et al., 2021; Rauf et al., 2023; Sohail et al., 2021; Zafar et al., 2021).

However, CNNs have the limitation of focusing only on local aspects of images, which means they fail to capture the global perspective of images (A. Khan et al., 2020). The small receptive fields of convolution filters (usually 3x3 or 5x5) allow them to learn local correlations in the images but no global-level information. Although many approaches have utilized dilated convolutions, large filters, and attention mechanisms to increase the receptive field, they still fall short in capturing the global perspective of images(Khalfaoui-Hassani et al., 2021; Yang et al., 2019). Due to the presence of intricate multi-level complex patterns dispersed globally within medical images, CNN-based systems that solely emphasize local patterns may demonstrate inadequate performance. The sample images from Lysto, NuClick and Lyon datasets are shown in Figure 1.

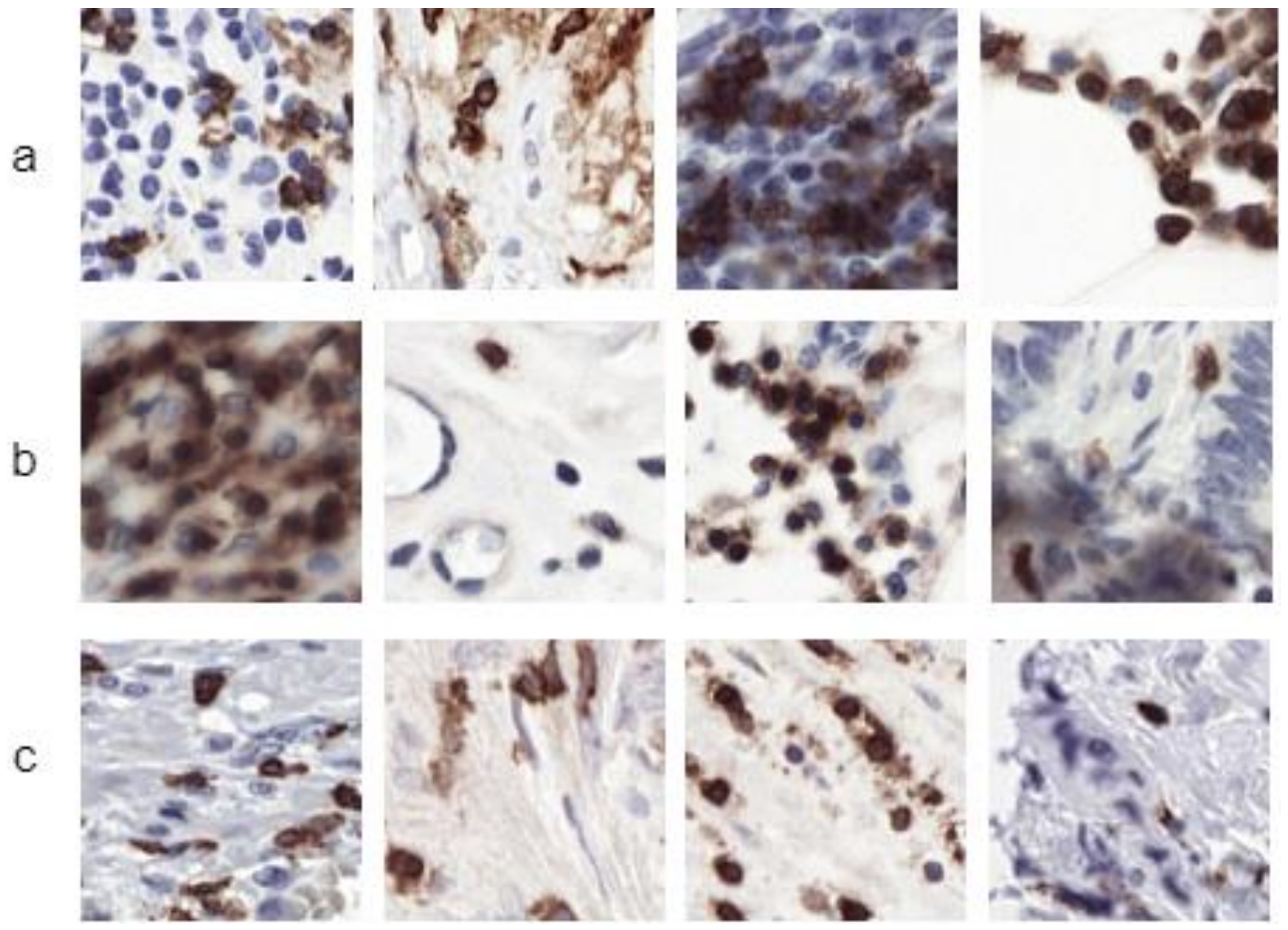

*Figure 1: Histopathological image taken from (a,b,c) Lysto, NuClick and Lyon datasets showing the presence of artifacts and lymphocytes.*

Recently, vision transformers have gained popularity due to their ability to model long-range dependencies in the images by utilizing their multi-head self-attention mechanism and positional embeddings (S. Khan et al., 2021). Various ViT-based approaches have shown promising results compared to CNNs (Shamshad et al., 2022). However, the poor image-related inductive bias of ViTs, high computation, and memory consumption due to multi-head attention, and fixed-sized image tokens may limit their performance for medical images (Y. Xu et al., 2021). In addition, ViT-based networks assume identical distributions for both training and test sets, however, in medical images, varying staining techniques, data acquisition methods, and scanners can introduce a domain shift, leading to poor generalization.

Therefore, in this study, we present a hybrid approach that leverages the benefits of both CNNs and ViTs for histopathological images. The proposed Channel Boosted Hybrid Vision Transformer network "CB-HVTNet" utilizes three main modules to learn highly correlated and domain-specific features, including a) the channel generator module, b) the channel exploitation module and c) the channel merging module. In the channel generation module we use three different channel generators to learn diverse channels from images, with each generator specialized in learning different types of features. Two of them are CNN-based to generate local-level information, and one is inspired by ViT architecture to capture global-level information. By combining the knowledge space of all these generators, we achieve a boosted channel space, which undergoes effective channel exploitation in the channel exploitation module and systematic channel fusion in the channel merging module, respectively. Later the proposed CB-HVTNet utilizes its region aware module and head to generate the final output.

Lymphocytes play a critical role in cancer analysis due to their high prognostic significance in assessing disease progression and the therapeutic efficacy of treatments such as chemotherapy or surgery (Orhan et al., 2020). However, their complex morphology, presence of noise, overlapping cells, and unclear boundaries make their manual analysis very challenging (X. Zhang et al., 2022). Therefore, an accurate automated diagnostic system for lymphocyte assessment can help pathologists in performing disease diagnosis and developing treatment plans. In this regard, the proposed approach “CB-HVT” enables a more reliable and efficient automated evaluation of lymphocytes, ultimately leading to better diagnosis and treatments for patients. Figure 2 shows the proposed framework's detailed workflow.

The significant contributions of the proposed CB-HVTNet are listed below:

- The proposed Channel Boosted Hybrid Vision Transformer Network “CB-HVTNet” integrates Convolutional Neural Networks (CNNs) with Vision Transformer-based (ViT) channel generators using a channel-boosting approach in the channel generation module. This approach effectively captures both local and global features, leading to more enriched feature representations and better learning outcomes.
- The channel merging module employs a novel fusion block to extract highly discriminant and domain-relevant features from multiple channel generators. This innovative approach contributes significantly to enhancing the accuracy and performance of the proposed method.
- The proposed technique has shown promising results on benchmark datasets, thereby providing evidence of its potential to be implemented as an effective diagnostic tool for lymphocyte assessment in histology images.

The remaining contents are arranged as follows: Related works are briefly summarised in Section 2. Section 3 presents the methodology of the developed CB-HVTNet in detail. We present the experimental results and discuss their implications in Section 4. Finally, in Section 6 we summarize the main findings and conclude the paper.

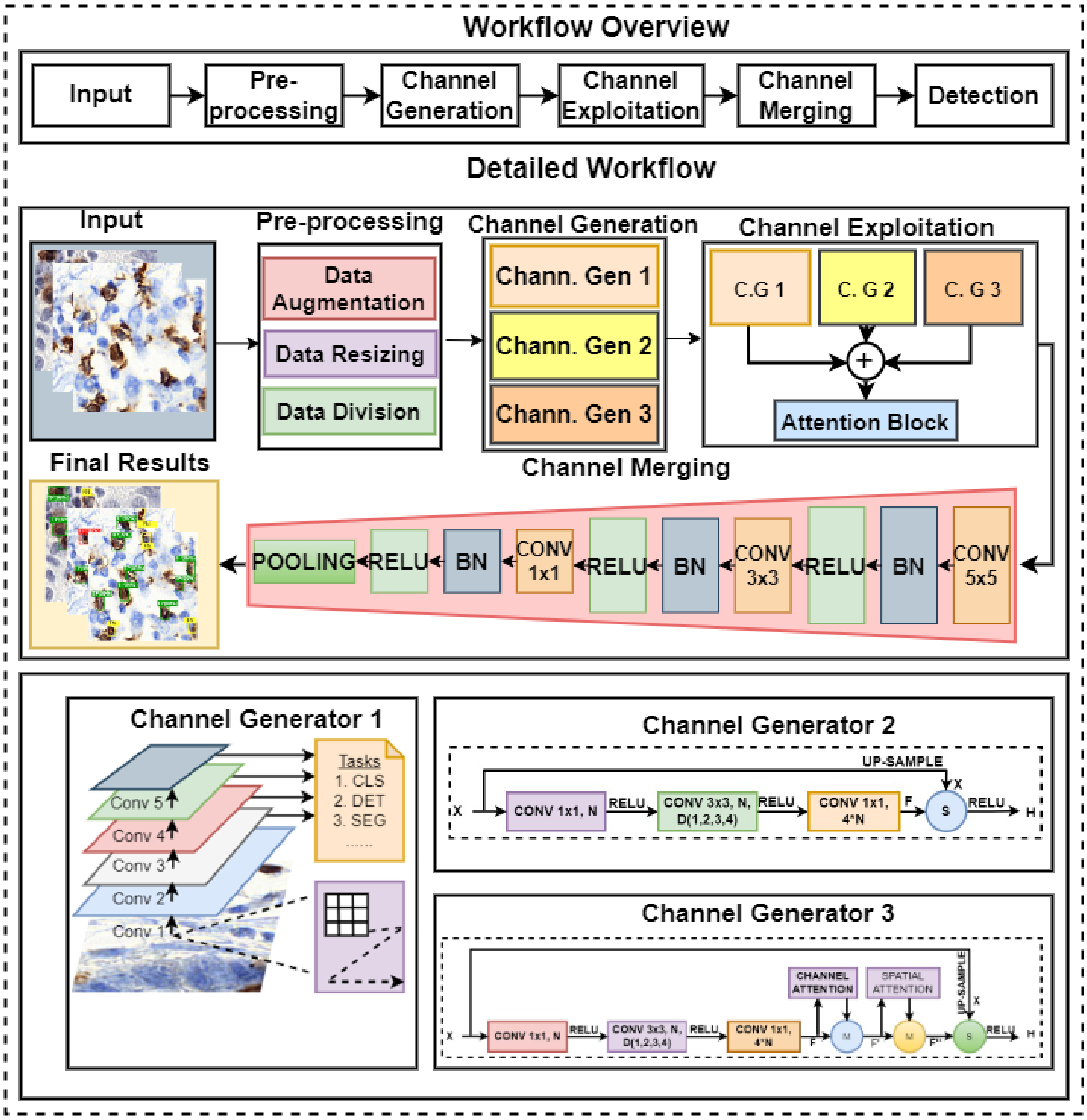

*Figure 2: Complete workflow of the proposed detection framework (CB-HVTNet)*

## 2. Related work

In the field of medical imaging, accurate object identification is of utmost importance for precise diagnosis and treatment. Various deep learning architectures have been developed for this purpose, with CNNs and transformers emerging as significant methods for object detection. Taking advantage of their ability to extract features from images and recognize patterns, CNNs, and transformers have been successfully used in medical imaging. Despite their different approaches to image data handling, both CNNs and transformers have shown great potential in medical image analysis.

### 2.1. CNN-based Methods

A considerable amount of research has been done using traditional image processing and machine learning-based algorithms for cell identification and detection in digital histopathology slides (Cohen, 2021; Sotiras et al., 2016; Wang et al., 2014) including region growth (Z. Xu et al., 2018) morphological operations (Malpica et al., 1997), and hand-crafted feature analysis (Madabhushi & Lee, 2016). However, recent developments in deep learning have revolutionized the field of computer vision, enabling algorithms to learn complex representations from raw data. As a result, deep learning algorithms have been applied to medical image analysis and have demonstrated the ability to perform at levels comparable to and, in some cases, surpassing those of human experts (Anwar et al., 2018; S. Graham, Chen, et al., 2019; S. Graham, Vu, et al., 2019).

Janowczyk et al. carried out a study to perform lymphocyte detection in histopathological images of cancer patients. In their work, they proposed a deep learning-based technique to classify lymphocytic and non-lymphocytic patches (Janowczyk & Madabhushi, 2016). Rijthoven et al. carried out lymphocyte detection in breast, colon, and prostate immunohistochemistry cancer images using a YOLO v3-based architecture (Van Rijthoven et al., n.d.). Linder et al. proposed a two-stage classification strategy to discriminate lymphocytic whole slide images. The first step involved the rough identification of a lymphocytic area, which was then subjected to the second-stage to obtain more accurate cell detection results (Linder et al., 2019). Swiderska-Chadaj et al. exploited FCN and U-Net to locate lymphocytes in IHC-stained images. In their study the performance of deep learning models in three types of lymphocytes containing regions, named artifact regular and clustered regions were compared (Swiderska-Chadaj et al., 2019).

Region-based CNNs (fast R-CNN, faster RCNN) (Li et al., 2019; Xue et al., 2021) have shown great performance in detecting complex objects, including lymphocytes (Özyurt, 2020; Sheng et al., 2020). These models exploit a smaller network as a region proposal network (RPN) to identify the probable regions that may contain the object. These selected regions are subsequently analyzed to determine the precise location and nature of the object. Zhang et al. exploited Mask R-CNN to develop a unified framework for panoptic segmentation in histology images (D. Zhang et al., 2018). Liu et al. carried out cell instance segmentation by incorporating a new module in the segmentation head of Mask R-CNN (Liu et al., 2019). They also proposed a feature map combination method to integrate the local and global level feature learning. Their findings from experiments demonstrated that integrating this additional module into the model enhanced the precision of cell instance segmentation. Kutlu et al. introduced a computer-aided automated approach that quickly identified and detected different types of WBC in blood images (Kutlu et al., 2020). Zafar et al. proposed a two-phase approach to identify tumor-infiltrating lymphocytes (TILs) in multiple cancer images(Zafar et al., 2021). Zhang et al. developed a novel architecture to evaluate TILs in hematoxylin and eosin-stained images of breast cancer (X. Zhang et al., 2022). Rauf et al. carried out lymphocyte detection using deep CNN-based technique. their method exploited two different architectures to effectively identify lymphocytes in histopathological images (Rauf et al., 2023).

## 2.2. Transformer-based Methods

Transformers, on the other hand, are a newer architecture and have shown significant contribution in object detection tasks specifically in medical imaging (Shamshad et al., 2023). Obeid et al. performed nucleus detection in histopathological images using a transformer (Obeid et al., 2022). Chen et al. employed transformers for gastric histopathological images and to extract features, they created a Global Information Module (GIM) and Local Information Module (LIM) which were CNN-based. Additionally, they incorporated the Inception-V3 architecture to acquire multi-scale local representations (H. Chen et al., 2022). Although ViT-based methods tackle the limitations of CNN, poor inductive bias and high computation make them unsuitable for many real-time problems, including medical diagnosis (Rehman & Khan, 2023). Recently, researchers have come up with the idea of merging both CNNs and transformers to get the benefits of both method (A. Khan et al., 2023). In this regard, Srinivas et al., introduced BoTNet where they modified the last three blocks of ResNet and significantly modified the ViT’s self-attention mechanism (Srinivas et al., 2021). Guo et al. introduced a transformer architecture in which they added pointwise and depthwise convolution before the self-attention module (Guo et al., 2021). Graham et al., replaced the patch embedding block with the convolutional stem to quickly classify images (B. Graham et al., 2021). Similarly, Chen et al, proposed the first hybrid transformer, TransUNet by combining transformer and UNet for medical image segmentation (J. Chen et al., 2021). Cao et al. introduced Swin-UNet in their work, in which they exploited the Swin attention module for medical image segmentation (Cao et al., 2021). Gao et al. utilized UNet architecture for medical image segmentation but replaced the last convolution block with the transformer block to incorporate the attention mechanism (Y. Gao et al., 2021).

Despite their effectiveness, the above described methods have their own limitations. A major concern is their high computational complexity, which can make them impractical for medical diagnosis in laboratories. Furthermore, some of these methods might not be able to fully capture the relevant information at each stage of the architecture, which could affect their overall performance.

## 3. Method

Automated assessment of lymphocytes in histology images is challenging due to the complex nature of tissue representation. Such complexity often leads to a high percentage of false positives, as well as difficulties in detecting lymphocytes that appear in clusters or exhibit different morphologies. In order to address these challenges, we have developed a novel framework for lymphocyte assessment. The detailed workflow of the proposed CB-HVTNet is depicted in Figure 3. CB-HVTNet consists of five main modules, named: a) channel generation module, b) channel exploitation module, c) channel merging module, d) region-aware module, and e) classification and detection head. Details of each module are elaborated upon in the following sections.

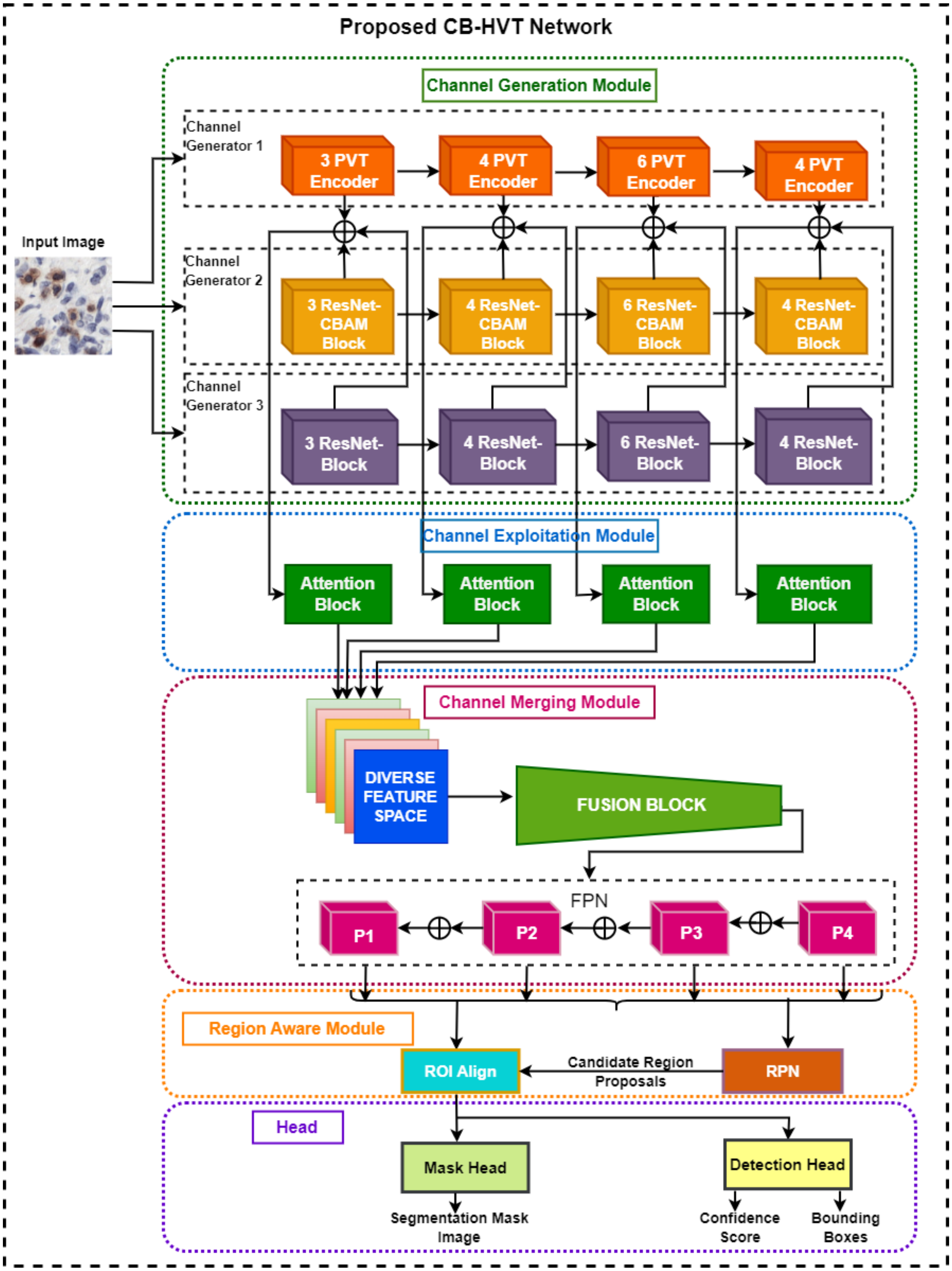

*Figure 3: Detailed workflow of the proposed architecture CB-HVTNet which has five main modules.*

## 3.1. Proposed Channel Boosted Hybrid Vision Transformer Network "CB-HVTNet"

The proposed CB-HVTNet begins by employing its channel generation module to generate boosted channels for a given input image, utilizing transfer learning to produce high-dimensional features. These boosted channels are then passed to the channel exploitation and channel merging modules for channel fusion and reduction, where domain-relevant and discriminant features are re-weighted to enhance their contributions. In the region-aware module, CB-HVTNet uses a Region Proposal Network (RPN) to extract objects containing probable regions. Finally, the classification and detection head generates the final output. Details of each module are described below.

### 3.1.1. Channel generation module

Given the complex patterns of medical images and high-level pattern variations at both tissue and cellular levels, we utilized the idea of channel boosting to generate diverse boosted channels. The proposed CB-HVTNet employes three heterogeneous architectures based on the concepts of vision transformer, spatial and channel attention, and residual connection to capture multi-level variations. The learned multi-variate feature space not only captured the global level context but also the local image representations. Our channel boosting approach involves domain adaptation-based transfer learning to extract channels from diverse architectures based on their unique learning abilities. To achieve this, we employed two pre-trained CNN-based networks and a transformer-based network.

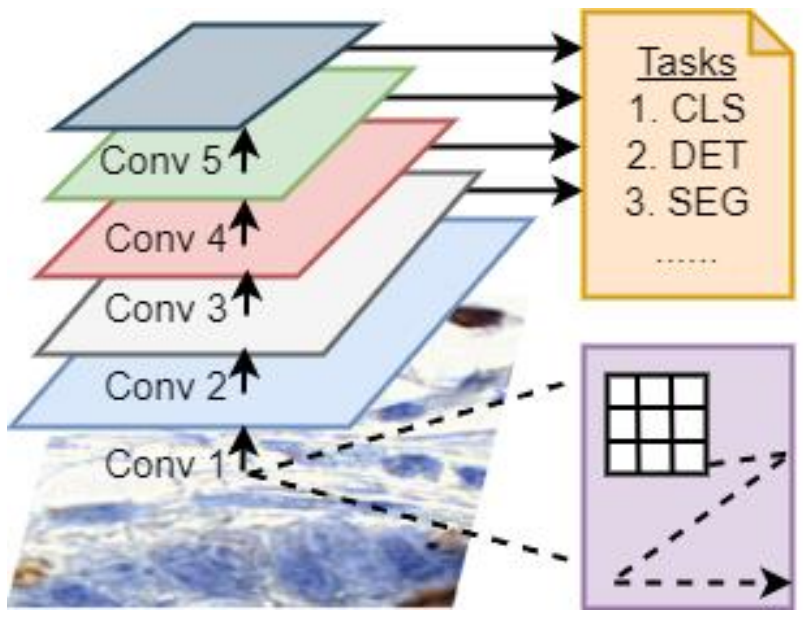

*Figure 4: Shrinking pyramid structure of PVT model*

For the transformer-based learner, we utilized Pyramid Vision Transformer (PVT) to learn long-range dependencies and global-level contextual information. In PVT, a group of convolutional layers initially processes the input image to extract low-level features. Then, the multi-scale transformer module processes these features to extract relevant information at different scales. The output of the transformer is then passed through a set of fully connected layers to produce the final output. The shrinking pyramid-like structure of PVT is shown in Figure 4.

The second channel generator in the CB-HVTNet is a pre-trained 50-layered Residual Network (ResNet-50), which has demonstrated excellent performance in various medical image tasks (Keren Evangeline et al., 2020). The main idea behind the residual connection is to enable the network to obtain the residual relationship between the input and output of a layer instead of directly learning the underlying mapping (Eq. 1) (K He et al., n.d.). It utilizes residual connections to enable reference-based learning and solve the problem of dead neurons.

$$Output = Input + F(Input) \quad (1)$$

The input is denoted by "$Input$", while "$F$" represents the residual function. The output of the residual block is computed by adding the input and the output of the residual function, which helps to create a shortcut connection between the input and output.

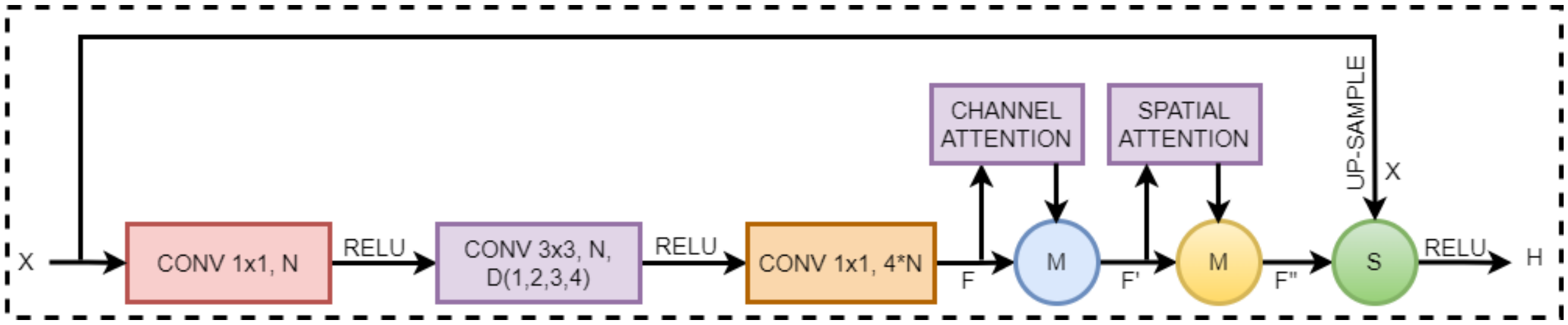

*Figure 5: Attention based channel generator which employs the idea of attention with residual connection*

Moreover, to capture class-specific features at both the channel and spatial levels, attention-based ResNet is employed. The architecture of the attention-based Channel Generator is shown in Figure 5. Eqs. 2 and 3 illustrate the concept of spatial and channel attention, respectively.

$$F' = M_C(F) \otimes F \tag{2}$$

$$F'' = M_S(F') \otimes F' \tag{3}$$

Here, the symbol ⊗ denotes element-wise multiplication, while $M_s(F)$ and $M_c(F)$ represent spatial attention and channel attention, respectively. The input feature map is represented by F. To refine the feature map, we perform an element-wise multiplication between the channel attention and the input feature map, resulting in a refined feature map F'. This refined feature map is then used in another element-wise multiplication with the spatial attention, resulting in the refined output feature map F". This process helps to enhance the feature map representation and improve the model's performance.

The utilization of these three diverse channel generators in HVT-CBNet resulted in the creation of a diverse and boosted feature space, which improves the ability to discriminate and identify objects with distinct boundaries, variable sizes, and shapes.

### 3.1.2. Channel exploitation module

The learned diverse and boosted channels from the channel generation module are exploited in the channel exploitation module to figure out the domain-relevant channels. The boosted channels from diverse learners are concatenated and weighted, using the attention mechanism (Eq. 2 and 3) to allow the network to focus on the most relevant channels while ignoring the ones with redundant information.

### 3.1.3. Channel merging module

The proposed approach utilizes a feature pyramid network in this module to extract feature maps at multiple levels of abstraction. This enables the model to capture high-level variations and semantic information present in the input images. Additionally, a novel feature merging block is employed to systematically reduce the channel dimension while retaining the most relevant features from the aggregated feature space. This block improves the accuracy and representation capacity of the proposed approach. Figure 6 illustrates the design of this feature merging block, which applies several sets of transformations, such as 3x3 and 1x1 convolutions, to the outputs of various channel generators.

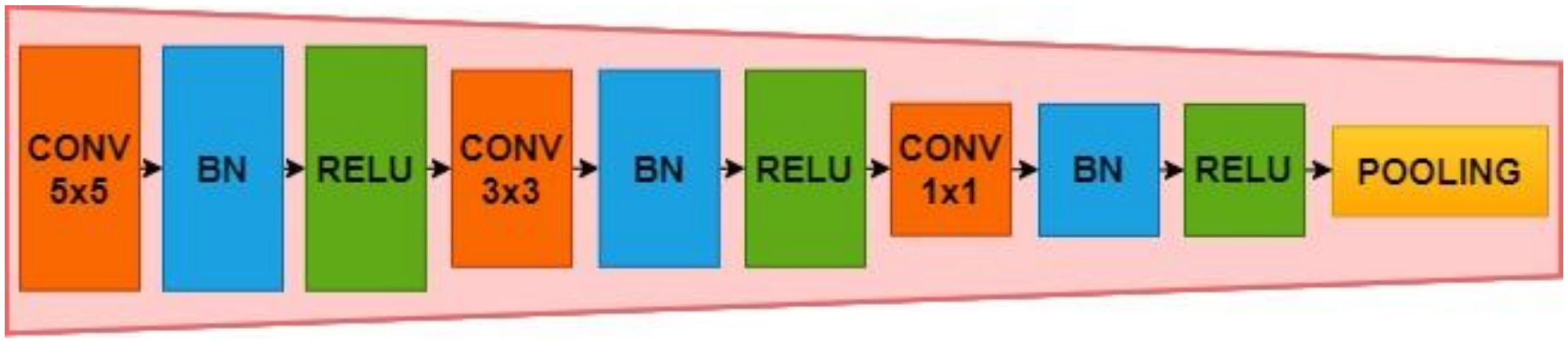

*Figure 6: Feature merging block used to merge the output of channel generators. The idea of bottleneck is employed in this merging block to obtain the most relevant features as output.*

### 3.1.4. Region aware module

The region-aware module of the proposed CB-HVT employs a Region Proposal Network (RPN) to identify probable regions that may contain lymphocytes (Ren et al., 2017). The selected region proposals are then passed to the ROI Align layer, which aligns each proposal region with its corresponding feature map. This allows for accurate sampling of a fixed-size feature map using bilinear interpolation, without compromising the spatial resolution of the original feature map.

### 3.1.5. Detection and segmentation head

The fixed-sized feature maps are fed to the detection and segmentation head of the proposed approach (Kaiming He et al., 2017). The detection head produces a set of bounding boxes for possible objects and their corresponding objectness scores, whereas the segmentation head produces a binary mask for each object, indicating its precise location in the image.

## 4.1. Error function of the proposed CB-HVTNet

The error function of the proposed approach is a combined loss for each of its detection and segmentation heads, given in Eq. 4, where $L_C$ is the Cross-Entropy Loss, $L_L$ is L1 Loss, and $L_B$ is the Binary Cross Entropy Loss.

$$L_{CB-HVI} = L_C + L_L + L_B \tag{4}$$

The detection head employs two loss functions, the Cross-Entropy and L1 loss for the prediction of the bounding box and class label, respectively (Eq. 5 and Eq. 6). The segmentation head of the proposed CB-HVTNet is the Binary Cross Entropy loss to predict the binary mask for each object in the image (Eq. 7).

$$L_C = -log(p_j[y_j]) \tag{5}$$

$$L_L = \frac{SUM|t_j - t * _j|}{N} \quad (6)$$

$$L_B = \frac{-1}{N} * sum\left(y_j * log(p_j) + (1 - y_j) * log(1 - p_j)\right) \quad (7)$$

The cross-entropy loss and L1 loss are represented in Eq. (5 & 6) $p_j$ in Eq. 5 is the likelihood and $j$ is the anchor and $y$ is the actual and predicted class labels. In Eq. 6 $t_j$ is the actual bounding box coordinates of the $j_{th}$ anchor while $t^*_j$ is the predicted bounding box coordinate. In Eq.7, shows binary cross-entropy loss and where $y_j$ is the true label and $p_j$ is the probability of the predicted class.

# 4. Experimental Results and Discussion

## 4.1. Datasets

In this study, the LYSTO and the NuClick datasets are utilized to assess the effectiveness of the proposed CB-HVTNet. The LYSTO dataset was released by the Lymphocyte Assessment and Hackathon (LYSTO) challenge organizers (Ciompi et al., 2019). It consists of 43 patients for breast colon and prostate cancer and released a total of 20k images. Among these, we selected 19 patients for training, 9 patients for validation, and 6 patients for testing. In contrast, the NuClick dataset, introduced by Koohbanani et al. (Alemi Koohbanani et al., 2020), comprises 471 images. Details of these datasets are depicted in Table 1.

*Table 1: Details of LYSTO and NuClick datasets are presented below*

| Dataset details | LYSTO dataset | NuClick dataset |
|---|---|---|
| **Image size** | 267 x 267 pixels | 256x256 pixels |
| **Staining** | Immunohistochemistry | Immunohistochemistry |
| **Train set** | 9000 | 471 |
| **Validation set** | 3000 | 99 |
| **Test set** | 3000 | 300 |

## 4.2. Training and implementation details

All the experiments of this work were conducted on an NVIDIA GTX machine with 1070 GPU and 8GB of memory, using an open-source PyTorch. Moreover, an open-source toolbox OpenMMLab was employed for the implementation of all the proposed and comparative models. The rest of the libraries and their versions are listed in Table 2, whereas, Table 3 lists the hyperparameters for all the models.

*Table 2: The details of libraries and their version.*

| Libraries | Versions |
|---|---|
| **Pytorch** | 1.12.1 |
| **Numpy** | 1.23.1 |
| **Opencv-python** | 4.6.0.66 |
| **CUDA version** | 11.8 |

*Table 3: Details of the selected hyperparameters for the proposed and comparative models.*

| Configuration | Value |
|---|---|
| **Epochs** | 30 |
| **Learning Rate** | 0.0025 |
| **Weight Decay** | 0.0001 |
| **Momentum** | 0.9 |
| **Optimizer** | SGD |
| **Batch size** | 4 |

## 4.3. Performance Metrics

F-score and recall were used as the performance metrics to analyze the performance of the proposed model and the comparison models. The F-score, which is the harmonic mean of precision and recall, is an unbiased and reliable measure, to evaluate the representation learning ability of the model especially when the data is imbalanced. The model's Recall indicates its ability to reliably identify true predictions. The formulas for F-score and recall are shown in Eq. 8 and Eq. 9, respectively.

$$F - score = \frac{2(Precision * Recall)}{Precision + Recall} \quad (8)$$

$$Recall = \frac{True\ Positive}{(True\ Positive + False\ Negative)} \quad (9)$$

## 4.4. Comparative study with existing methods

For a fair comparison, we compared our proposed CB-HVTNet with other existing models for lymphocyte assessment. In this regard, we selected MaskRCNN as the state-of-the-art two-stage detector, YOLO and SC-Net as the latest single-stage detectors, and Unet as a semantic segmentation model. We evaluated our proposed CB-HVTNet against these existing models on the test sets of LYSTO and NuClick datasets to determine its effectiveness. We utilized the evaluation metrics discussed in Section 4.3 to evaluate the performance of these models comprehensively. The results of the proposed model with other comparison architectures are shown in Figure 7 and Figure 8.

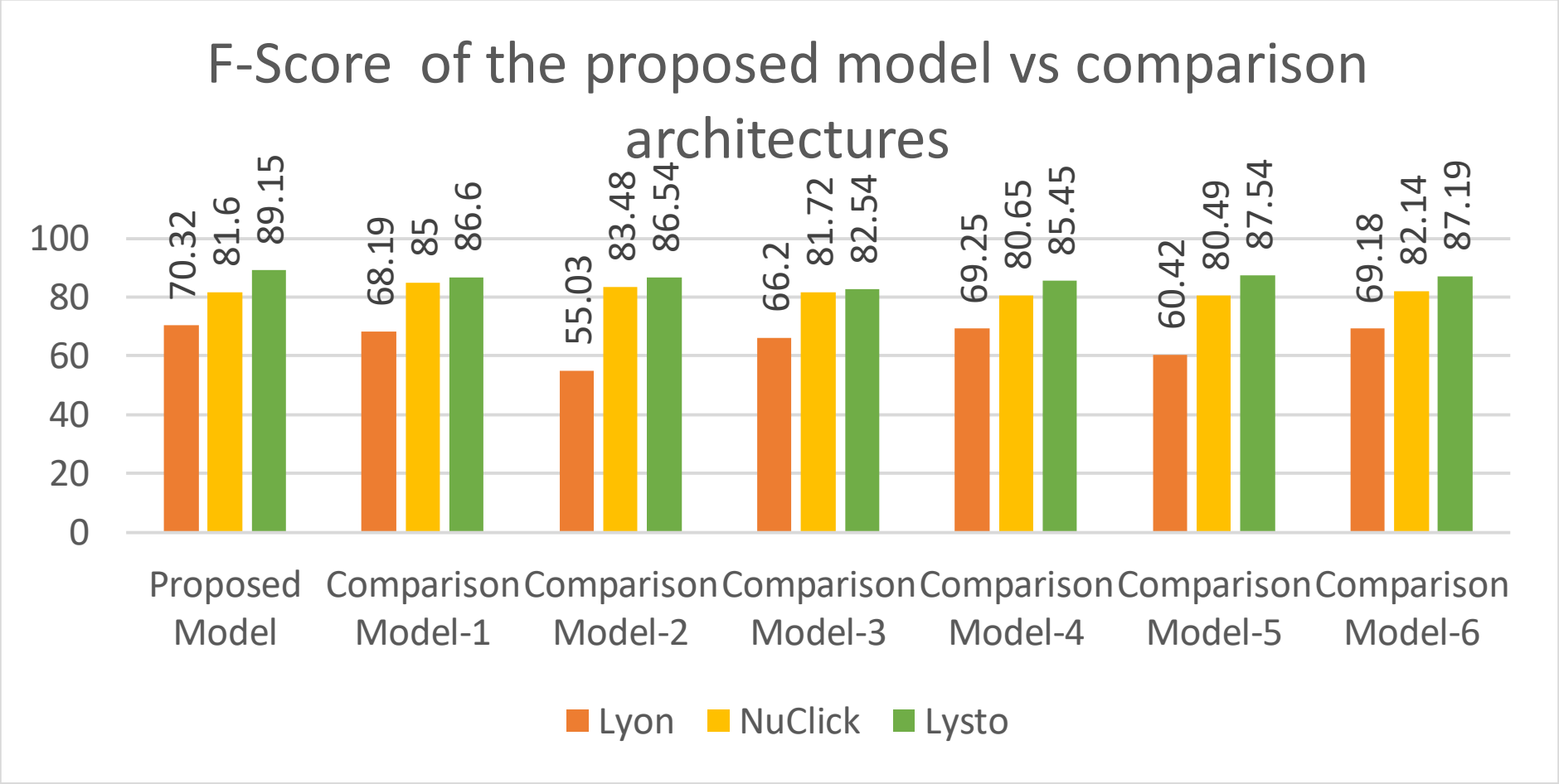

*Figure 7: F-score of the proposed model on Lysto, NuClick and Lyon (Test set) datasets.*

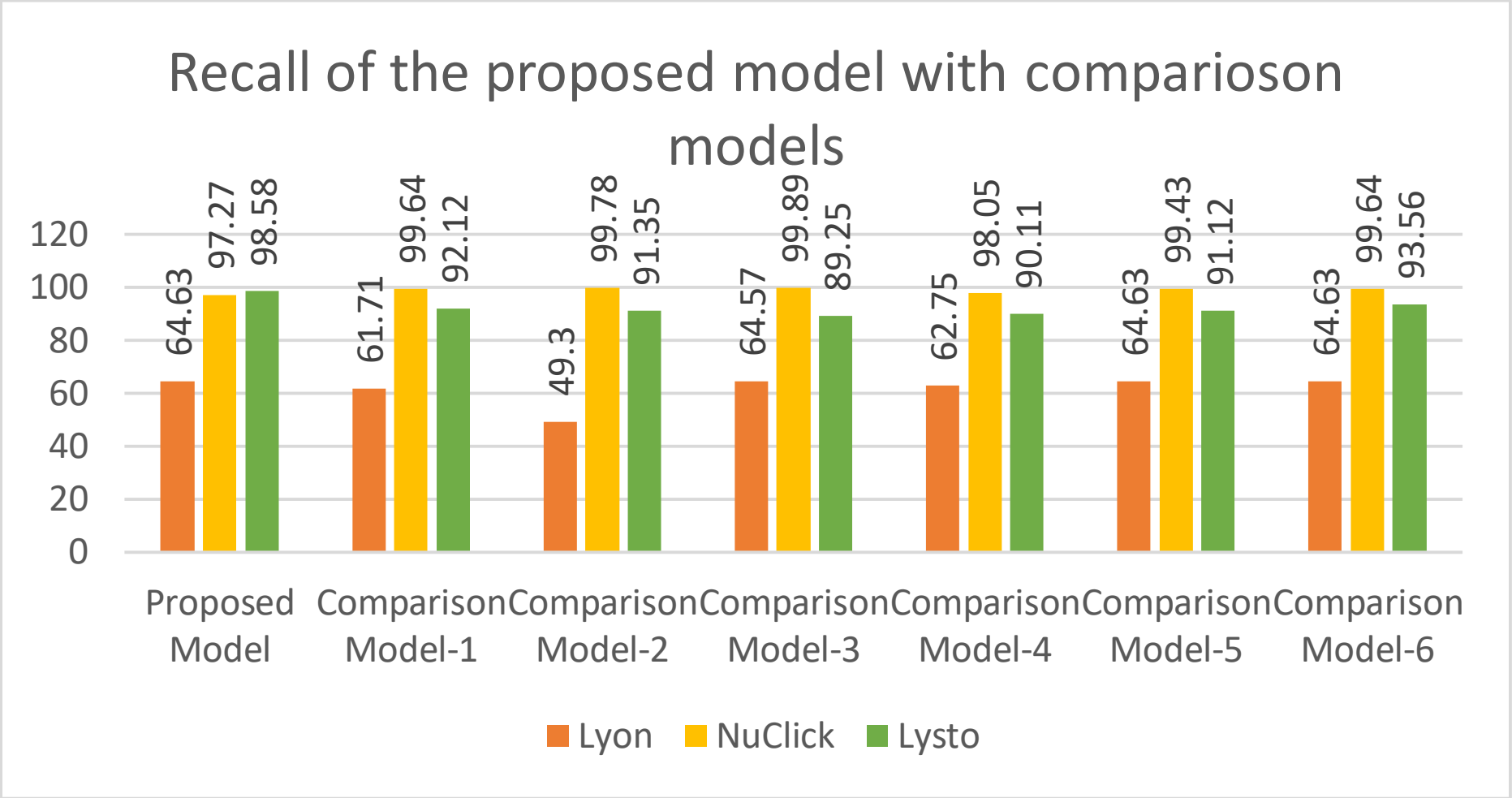

*Figure 8: Recall of the proposed model on Lysto, NuClick and Lyon (Test set) datasets.*

### 4.5. Quantitative results of CB-HVTNet and the comparative models

The quantitative results of the proposed CB-HVTNet and the comparative models are shown in Table 2. These models are evaluated based on F-score and recall. The proposed CB-HVTNet outperformed Mask RCNN, SC-Net, and YOLO with an F-score of 0.88 and 0.82 on LYSTO and Nuclick datasets, respectively. In terms of F-score, SC-Net performed well on both datasets as compared to Mask RCNN and YOLO. SC-Net detected lymphocytes better than YOLO with a recall of 0.82 on the LYSTO dataset, whereas YOLO showed an improved recall when evaluated on Nuclick dataset (Figure 8). The results are presented in Table 4.

*Table 4: Comparison of the proposed CB-HVNet on Lysto and NuClick with state-of-the-art models*

| **Technique** | **Lysto** | | **Nuclick** | |
|---|---|---|---|---|
| | F-score | Recall | F-score | Recall |
| **Proposed CB-HVNet** | 0.88 | 0.93 | 0.82 | 0.99 |
| **Mask RCNN** | 0.80 | 0.73 | 0.78 | 0.67 |
| **SC-Net** | 0.85 | 0.82 | 0.80 | 0.84 |
| **YOLO** | 0.80 | 0.69 | 0.76 | 0.86 |

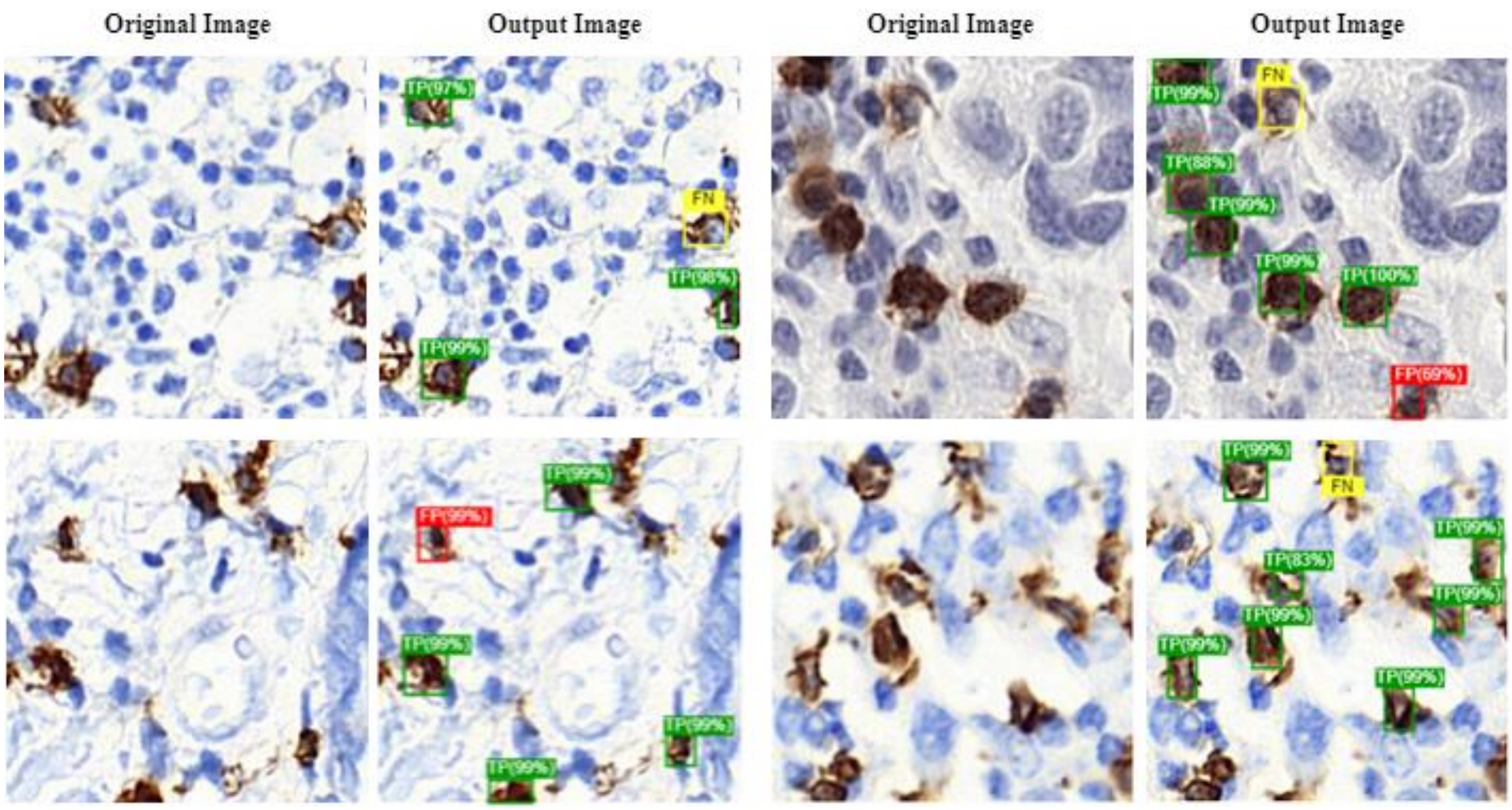

*Figure 9: Results of the proposed CB-HVTNet on Lysto and NuClick datasets. (Legend: green: TP, red: FP, yellow: FN)*

## 4.6. Qualitative results of CB-HVNet and the comparative models

To further understand the effectiveness of the proposed CB- HVTNet, a qualitative analysis was conducted in addition to the quantitative evaluation. The outcomes of our model while evaluating lymphocytes in the LYSTO and NuClick datasets are shown in Figure 9. The findings clearly show that our proposed CB-HVTNet performs better not only in terms of quantitative evaluation but also in terms of visual interpretation.

## 4.7. Ablation study

We carried out extensive experiments to assess the efficacy of our proposed architecture design. Specifically, we performed various experiments to identify optimal combination of channel generators in the channel generation module and the fusion block in the channel merging module.

### 4.7.1. Channel generation module

This section describes the channel generation module, which consists of several channel generators used to generate boosted channels. We experimented with different combinations of architectures to identify the best-performing model. Specifically, we employed ResNet50, ResNet50 with CBAM, Autoencoder, and PVT backbone, with various combinations. Table 5 summarizes the different architecture combinations used for channel generation.

*Table 5: Architectural details for different channel generators*

| Model | Description |
|---|---|
| **Channel Generator-1** | ResNet, PVT |
| **Channel Generator-2** | ResNet-50, ResNet-CBAM, PVT, Convolutional Autoencoder |
| **Channel Generator-3** | ResNet-50, ResNet-CBAM, ConvAutoencoder |
| **Channel Generator-4** | ResNet-50, ResNet-CBAM, PVT, ResNet-101 |
| **Channel Generator-5** | ResNet-50, ResNet-CBAM, ResNext, ResNet-101 |
| **Channel Generator-6** | ResNet-CBAM, ResNext |

### 4.7.2. Channel merging module

Experiments with different combinations of fusion blocks are done to identify the most optimal method for merging the generated diverse boosted channels. Table 6 presents the architecture of these fusion blocks, and Table 7 reports their results in terms of F-score and recall. Results show that the Channel Merger-6 when used with Channel Generator-6 performed better than the other combinations but the proposed CB-HVTNet performed better in comparison with all six in-house built comparison architectures.

*Table 6: Architectural details of several fusion blocks utilized in feature merger module*

| Model | Fusion Block Architectures |
| --- | --- |
| **Channel Merger-1** | 5x5 conv, Batch Norm, ReLU, 3x3 conv, Batch Norm, ReLU, 1x1 conv, Batch Norm, ReLU, Pooling |
| **Channel Merger-2** | 5x5 conv, Batch Norm, ReLU, 3x3 conv, Batch Norm, ReLU, 1x1 conv, Batch Norm, ReLU, Pooling |
| **Channel Merger-3** | 5x5 conv, Batch Norm, ReLU, 3x3 conv, Batch Norm, ReLU, 1x1 conv, Batch Norm, ReLU, Pooling |
| **Channel Merger-4** | 7x7 conv, Batch Norm, ReLU, 5x5 conv, Batch Norm, ReLU, 1x1 conv, Batch Norm, ReLU, Pooling |
| **Channel Merger-5** | 3x3 conv, Batch Norm, ReLU, 3x3 conv, Batch Norm, ReLU, 1x1 conv, Batch Norm, ReLU, Pooling |
| **Channel Merger-6** | 5x5 conv, Batch Norm, ReLU, 3x3 conv, Batch Norm, ReLU, 1x1 conv, Batch Norm, ReLU, Pooling |

*Table 7: Comparison of various settings for the proposed "CB-HVTNet"*

| | **LYSTO** | | **NuClick** | |
| --- | --- | --- | --- | --- |
| **Backbone** | F-Score | Recall | F-Score | Recall |
| **Comparison Model-1 (Channel Generator-1 + Channel Merger-1)** | 86.60 | 92.12 | 85.00 | 99.78 |
| **Comparison Model-2 (Channel Generator-2 + Channel Merger-2)** | 86.54 | 91.35 | 83.48 | 99.89 |
| **Comparison Model-3 (Channel Generator-3 + Channel Merger-3)** | 82.54 | 89.25 | 81.72 | 98.05 |
| **Comparison Model-4 (Channel Generator-4 + Channel Merger-4)** | 85.45 | 90.11 | 80.65 | 99.43 |
| **Comparison Model-5 (Channel Generator-5 + Channel Merger-5)** | 87.54 | 91.12 | 80.49 | 99.64 |
| **Comparison Model-6 (Channel Generator-6 + Channel Merger-6)** | 88.19 | 93.56 | 82.14 | 99.64 |

### 4.8. Generalization analysis

We also evaluated the performance of our proposed CB-HVTNet on an unseen test set, which was released as part of the LYON'19 challenge. The challenge organizers did not provide any training set for the test images. However, we were able to evaluate our pre-trained model on the provided 441 ROI images. While the test labels were kept hidden by the challenge organizers, we obtained the results from the leader board. The results indicate that the proposed CB-HVTNet achieved an F-score of 0.70 and a recall of 0.64 in identifying lymphocytes in these images (Figures 7 and 8).

## 5. Conclusion

Accurate and automated lymphocyte assessment plays an important role in cancer analysis, offering significant prognostic value. In this work, a Channel Boosted Hybrid Vision Transformer "CB-HVT" is proposed to identify lymphocytes in IHC-stained histology images. In the proposed approach, channel generation, channel exploitation, and channel merging modules are utilized to increase the model's representation learning ability. The channel generator module of CB-HVT exploits both CNN-based and transformer-based architectures to generate informative channels, enabling robust feature extraction. These channels are then systematically merged and fused using a novel channel merger module, facilitating the integration of complementary information. The proposed CB-HVTNet identified lymphocytes even in the existence of significant intra-class resemblance, occlusion, and artifacts due to the flaws in the lens and blurring effects. The proposed technique also uses a region-aware module to recognize probable lymphocytic areas and send them to the detection and segmentation module for additional examination. We evaluated the performance of CB-HVTNet by comparing it with various state-of-the-art models in lymphocyte detection and segmentation. The results indicate that the proposed approach outperforms other existing models in terms of F-score and recall. The effectiveness of CB-HVTNet's channel generation and merging, as well as its use of channel boosting and transfer learning, contributes to its superior performance. The proposed CB-HVTNet holds great potential for enhancing the efficacy and accuracy of cancer diagnosis and treatment planning through improved lymphocyte identification and analysis. This contribution advances the field of medical image analysis and paves the way for more effective cancer management.

## Acknowledgments

This work has been conducted at the pattern recognition lab, Pakistan Institute of Engineering and Applied Sciences, Islamabad, Pakistan. We acknowledge Pakistan Institute of Engineering and Applied Sciences (PIEAS) for a healthy research environment which led to the research work presented in this article.

## Authors Contributions

Momina Liaqat Ali: Conceptualization, Methodology, Software, Visualization, Validation, Writing – writing, review & editing

Zunaira Rauf: Conceptualization, Writing - review & editing.

Asifullah Khan: Conceptualization, Writing - review & editing, Supervision, Resources.

Anabia Sohail: Formal analysis, Review & editing

Rafi Ullah: Review & editing.

Jeonghwan Gwak: Review & editing.

## Competing interests

The authors declare no competing financial and/or non-financial interests about the described work.

## Additional information

Correspondence and requests for materials should be addressed to A.K.

## References

Alemi Koohbanani, N., Jahanifar, M., Zamani Tajadin, N., & Rajpoot, N. (2020). NuClick: A deep learning framework for interactive segmentation of microscopic images. *Medical Image Analysis*, *65*. https://doi.org/10.1016/j.media.2020.101771

Anwar, S. M., Majid, M., Qayyum, A., Awais, M., Alnowami, M., & Khan, M. K. (2018). Medical Image Analysis using Convolutional Neural Networks: A Review. In *Journal of Medical Systems*. https://doi.org/10.1007/s10916-018-1088-1

Cao, H., Wang, Y., Chen, J., Jiang, D., Zhang, X., Tian, Q., & Wang, M. (2021). *Swin-Unet: Unet-like Pure Transformer for Medical Image Segmentation*. 205–218. https://doi.org/10.1007/978-3-031-25066-8_9

Chen, H., Li, C., Wang, G., Li, X., Mamunur Rahaman, M., Sun, H., Hu, W., Li, Y., Liu, W., Sun, C., Ai, S., & Grzegorzek, M. (2022). GasHis-Transformer: A multi-scale visual transformer approach for gastric histopathological image detection. *Pattern Recognition*, *130*, 108827. https://doi.org/10.1016/J.PATCOG.2022.108827

Chen, J., Lu, Y., Yu, Q., Luo, X., Adeli, E., Wang, Y., Lu, L., Yuille, A. L., & Zhou, Y. (2021). *TransUNet: Transformers Make Strong Encoders for Medical Image Segmentation*. https://arxiv.org/abs/2102.04306v1

Ciompi, F., Jiao, Y., & Laak, J. van der. (2019). *Lymphocyte Assessment Hackathon (LYSTO)*. https://doi.org/10.5281/ZENODO.3513571

Cohen, S. (2021). The evolution of machine learning: past, present, and future. *Artificial Intelligence and Deep Learning in Pathology*, 1–12. https://doi.org/10.1016/B978-0-323-67538-3.00001-4

Gao, J., Jiang, Q., Zhou, B., Chen, D., Gao, J., Jiang, Q., Zhou, B., & Chen, D. (2019). Convolutional neural networks for computer-aided detection or diagnosis in medical image analysis: An overview. *Mathematical Biosciences and Engineering 2019 6:6536*, *16*(6), 6536–6561. https://doi.org/10.3934/MBE.2019326

Gao, Y., Zhou, M., & Metaxas, D. N. (2021). UTNet: A Hybrid Transformer Architecture for Medical Image Segmentation. *Lecture Notes in Computer Science (Including Subseries Lecture Notes in Artificial Intelligence and Lecture Notes in Bioinformatics)*, *12903 LNCS*, 61–71. https://doi.org/10.1007/978-3-030-87199-4_6/COVER

Graham, B., El-Nouby, A., Touvron, H., Stock, P., Joulin, A., Jégou, H., & Douze, M. (2021). LeViT: a Vision Transformer in ConvNet's Clothing for Faster Inference. *Proceedings of the IEEE International Conference on Computer Vision*, 12239–12249. https://arxiv.org/abs/2104.01136v2

Graham, S., Chen, H., Gamper, J., Dou, Q., Heng, P. A., Snead, D., Tsang, Y. W.,

& Rajpoot, N. (2019). MILD-Net: Minimal information loss dilated network for gland instance segmentation in colon histology images. *Medical Image Analysis*. https://doi.org/10.1016/j.media.2018.12.001

Graham, S., Vu, Q. D., Raza, S. E. A., Azam, A., Tsang, Y. W., Kwak, J. T., & Rajpoot, N. (2019). Hover-Net: Simultaneous segmentation and classification of nuclei in multi-tissue histology images. *Medical Image Analysis*, *58*, 1–18. https://doi.org/10.1016/j.media.2019.101563

Guo, J., Han, K., Wu, H., Tang, Y., Chen, X., Wang, Y., & Xu, C. (2021). CMT: Convolutional Neural Networks Meet Vision Transformers. *Proceedings of the IEEE Computer Society Conference on Computer Vision and Pattern Recognition*, *2022-June*, 12165–12175. https://doi.org/10.1109/CVPR52688.2022.01186

He, K, Zhang, X., Ren, S., recognition, J. S. pattern, & 2016, undefined. (n.d.). Deep residual learning for image recognition. *Openaccess.Thecvf.Com*. Retrieved September 22, 2022, from http://openaccess.thecvf.com/content_cvpr_2016/html/He_Deep_Residual_Learning_CVPR_2016_paper.html

He, Kaiming, Gkioxari, G., Dollár, P., & Girshick, R. (2017). Mask R-CNN. *IEEE Transactions on Pattern Analysis and Machine Intelligence*, *42*(2), 386–397. https://doi.org/10.1109/TPAMI.2018.2844175

Janowczyk, A., & Madabhushi, A. (2016). Deep learning for digital pathology image analysis: A comprehensive tutorial with selected use cases. *Journal of Pathology Informatics*, *7*(1). https://doi.org/10.4103/2153-3539.186902

Ke, J., Lu, Y., Shen, Y., Zhu, J., Zhou, Y., Huang, J., Yao, J., Liang, X., Guo, Y., Wei, Z., Liu, S., Huang, Q., Jiang, F., & Shen, D. (2023). ClusterSeg: A crowd cluster pinpointed nucleus segmentation framework with cross-modality datasets. *Medical Image Analysis*, *85*, 102758. https://doi.org/10.1016/J.MEDIA.2023.102758

Keren Evangeline, I., Glory Precious, J., Pazhanivel, N., & Angeline Kirubha, S. P. (2020). Automatic Detection and Counting of Lymphocytes from Immunohistochemistry Cancer Images Using Deep Learning. *Journal of Medical and Biological Engineering*, *40*(5), 735–747. https://doi.org/10.1007/s40846-020-00545-4

Khalfaoui-Hassani, I., Pellegrini, T., & Masquelier, T. (2021). *Dilated convolution with learnable spacings*. https://arxiv.org/abs/2112.03740v4

Khan, A., Rauf, Z., Sohail, A., Rehman, A., Asif, H., Asif, A., & Farooq, U. (2023). A survey of the Vision Transformers and its CNN-Transformer based Variants. *ArXiv Preprint ArXiv:2305.09880*.

Khan, A., Sohail, A., Zahoora, U., & Qureshi, A. S. (2020). A survey of the recent architectures of deep convolutional neural networks. *Artificial Intelligence*

*Review*, *53*(8), 5455–5516. https://doi.org/10.1007/s10462-020-09825-6

Khan, S. H., Sohail, A., Khan, A., Hassan, M., Lee, Y. S., Alam, J., Basit, A., & Zubair, S. (2021). COVID-19 detection in chest X-ray images using deep boosted hybrid learning. *Computers in Biology and Medicine*, *137*, 104816. https://doi.org/10.1016/J.COMPBIOMED.2021.104816

Khan, S., Naseer, M., Hayat, M., Zamir, S. W., Khan, F. S., & Shah, M. (2021). Transformers in Vision: A Survey. *ACM Computing Surveys*, *54*(10). https://doi.org/10.1145/3505244

Kutlu, H., Avci, E., & Özyurt, F. (2020). White blood cells detection and classification based on regional convolutional neural networks. *Medical Hypotheses*, *135*, 109472. https://doi.org/10.1016/J.MEHY.2019.109472

Li, Z., Dong, M., Wen, S., Hu, X., Zhou, P., & Zeng, Z. (2019). CLU-CNNs: Object detection for medical images. *Neurocomputing*, *350*, 53–59. https://doi.org/10.1016/j.neucom.2019.04.028

Linder, N., Taylor, J. C., Colling, R., Pell, R., Alveyn, E., Joseph, J., Protheroe, A., Lundin, M., Lundin, J., & Verrill, C. (2019). Deep learning for detecting tumour-infiltrating lymphocytes in testicular germ cell tumours. *Journal of Clinical Pathology*, *72*(2), 157–164. https://doi.org/10.1136/jclinpath-2018-205328

Liu, D., Zhang, D., Song, Y., Zhang, C., Zhang, F., O'Donnell, L., & Cai, W.

(2019). Nuclei segmentation via a deep panoptic model with semantic feature fusion. *IJCAI International Joint Conference on Artificial Intelligence*, *2019-August*, 861–868. https://doi.org/10.24963/IJCAI.2019/121

Madabhushi, A., & Lee, G. (2016). Image analysis and machine learning in digital pathology: Challenges and opportunities. *Medical Image Analysis*, *33*, 170–175. https://doi.org/10.1016/J.MEDIA.2016.06.037

Malpica, N., Solórzano, C. O. de, Vaquero, J. J., Santos, A., Vallcorba, I., García-Sagredo, J. M., & Pozo, F. del. (1997). Applying watershed algorithms to the segmentation of clustered nuclei. *Cytometry*, *28*(4), 289–297. https://doi.org/10.1002/(SICI)1097-0320(19970801)28:4<289::AID-CYTO3>3.0.CO;2-7

Obeid, A., Mahbub, T., Javed, S., Dias, J., & Werghi, N. (2022). NucDETR: End-to-End Transformer for Nucleus Detection in Histopathology Images. *Lecture Notes in Computer Science (Including Subseries Lecture Notes in Artificial Intelligence and Lecture Notes in Bioinformatics)*, *13574 LNCS*, 47–57. https://doi.org/10.1007/978-3-031-17266-3_5/COVER

Orhan, A., Vogelsang, R. P., Andersen, M. B., Madsen, M. T., Hölmich, E. R., Raskov, H., & Gögenur, I. (2020). The prognostic value of tumour-infiltrating lymphocytes in pancreatic cancer: a systematic review and meta-analysis. *European Journal of Cancer*, *132*, 71–84.

https://doi.org/10.1016/J.EJCA.2020.03.013

Özyurt, F. (2020). A fused CNN model for WBC detection with MRMR feature selection and extreme learning machine. *Soft Computing*, *24*(11), 8163–8172. https://doi.org/10.1007/S00500-019-04383-8/TABLES/5

Rauf, Z., Sohail, A., Khan, S. H., Khan, A., Gwak, J., & Maqbool, M. (2023). Attention-guided multi-scale deep object detection framework for lymphocyte analysis in IHC histological images. *Microscopy (Oxford, England)*, *72*(1), 27–42. https://doi.org/10.1093/jmicro/dfac051

Rehman, A., & Khan, A. (2023). MaxViT-UNet: Multi-Axis Attention for Medical Image Segmentation. *ArXiv Preprint ArXiv:2305.08396*. https://arxiv.org/abs/2305.08396v2

Ren, S., He, K., Girshick, R., & Sun, J. (2017). Faster R-CNN: Towards Real-Time Object Detection with Region Proposal Networks. *IEEE Transactions on Pattern Analysis and Machine Intelligence*, *39*(6), 1137–1149. https://doi.org/10.1109/TPAMI.2016.2577031

Shamshad, F., Khan, S., Zamir, S. W., Khan, M. H., Hayat, M., Khan, F. S., & Fu, H. (2022). *Transformers in Medical Imaging: A Survey*. https://arxiv.org/abs/2201.09873v1

Shamshad, F., Khan, S., Zamir, S. W., Khan, M. H., Hayat, M., Khan, F. S., & Fu, H. (2023). Transformers in medical imaging: A survey. *Medical Image*

*Analysis*, 102802. https://doi.org/10.1016/j.media.2023.102802

Sheng, B., Zhou, M., Hu, M., Li, Q., Sun, L., & Wen, Y. (2020). A blood cell dataset for lymphoma classification using faster R-CNN. *Http://Mc.Manuscriptcentral.Com/Tbeq*, *34*(1), 413–420. https://doi.org/10.1080/13102818.2020.1765871

Sohail, A., Khan, A., Nisar, H., Tabassum, S., & Zameer, A. (2021). Mitotic nuclei analysis in breast cancer histopathology images using deep ensemble classifier. *Medical Image Analysis*, *72*, 102121.

Sotiras, A., Gaonkar, B., Eavani, H., Honnorat, N., Varol, E., Dong, A., & Davatzikos, C. (2016). Machine learning as a means toward precision diagnostics and prognostics. *Machine Learning and Medical Imaging*, 299–334. https://doi.org/10.1016/B978-0-12-804076-8.00010-4

Srinivas, A., Lin, T. Y., Parmar, N., Shlens, J., Abbeel, P., & Vaswani, A. (2021). Bottleneck Transformers for Visual Recognition. *Proceedings of the IEEE Computer Society Conference on Computer Vision and Pattern Recognition*, 16514–16524. https://doi.org/10.1109/CVPR46437.2021.01625

Swiderska-Chadaj, Z., Pinckaers, H., van Rijthoven, M., Balkenhol, M., Melnikova, M., Geessink, O., Manson, Q., Sherman, M., Polonia, A., Parry, J., Abubakar, M., Litjens, G., van der Laak, J., & Ciompi, F. (2019). Learning to detect lymphocytes in immunohistochemistry with deep learning. *Medical*

*Image Analysis*. https://doi.org/10.1016/j.media.2019.101547

Van Rijthoven, M., Swiderska-Chadaj, Z., Seeliger, K., Van Der Laak, J., & Ciompi, F. (n.d.). *You Only Look on Lymphocytes Once*.

Wang, H., Cruz-Roa, A., Basavanhally, A., Gilmore, H., Shih, N., Feldman, M., Tomaszewski, J., Gonzalez, F., & Madabhushi, A. (2014). Mitosis detection in breast cancer pathology images by combining handcrafted and convolutional neural network features. *Journal of Medical Imaging*, *1*(3), 034003. https://doi.org/10.1117/1.JMI.1.3.034003

Xu, Y., Zhang, Q., Zhang, J., & Tao, D. (2021). ViTAE: Vision Transformer Advanced by Exploring Intrinsic Inductive Bias. *Advances in Neural Information Processing Systems*, *34*(NeurIPS), 28522–28535.

Xu, Z., Fernádez Moro, C., Kuznyecov, D., Bozóky, B., Dong, L., & Zhang, Q. (2018). Tissue region growing for hispathology image segmentation. *ACM International Conference Proceeding Series*, 86–92. https://doi.org/10.1145/3288200.3288213

Xue, R., Xiang, W., & Deng, Y. (2021). Improved Faster R-CNN Based on CSP-DPN. *Procedia Computer Science*, *199*, 1490–1497. https://doi.org/10.1016/j.procs.2022.01.190

Yang, L., Song, Q., Wu, Y., & Hu, M. (2019). Attention Inspiring Receptive-Fields Network for Learning Invariant Representations. *IEEE Transactions on*

*Neural Networks and Learning Systems*, *30*(6), 1744–1755. https://doi.org/10.1109/TNNLS.2018.2873722

Zafar, M. M., Rauf, Z., Sohail, A., Khan, A. R., Obaidullah, M., Khan, S. H., Lee, Y. S., & Khan, A. (2021). Detection of Tumour Infiltrating Lymphocytes in CD3 and CD8 Stained Histopathological Images using a Two-Phase Deep CNN. *Photodiagnosis and Photodynamic Therapy*, *37*(September 2021), 102676. https://doi.org/10.1016/j.pdpdt.2021.102676

Zhang, D., Song, Y., Liu, D., Jia, H., Liu, S., Xia, Y., Huang, H., & Cai, W. (2018). Panoptic segmentation with an end-to-end cell R-CNN for pathology image analysis. *Lecture Notes in Computer Science (Including Subseries Lecture Notes in Artificial Intelligence and Lecture Notes in Bioinformatics)*, *11071 LNCS*, 237–244. https://doi.org/10.1007/978-3-030-00934-2_27/COVER

Zhang, X., Zhu, X., Tang, K., Zhao, Y., Lu, Z., & Feng, Q. (2022). DDTNet: A dense dual-task network for tumor-infiltrating lymphocyte detection and segmentation in histopathological images of breast cancer. *Medical Image Analysis*. https://doi.org/10.1016/j.media.2022.102415